\documentclass[%
 reprint,
 amsmath,amssymb,
 aps,
]{revtex4-1}

\usepackage{graphicx}
\usepackage{dcolumn}
\usepackage{bm}
\usepackage{subfigure}
\usepackage{rotating}
\usepackage{float}
\usepackage{verbatim}
\usepackage{longtable}
\usepackage{txfonts}
\usepackage{color,soul}
\usepackage{epstopdf}
\usepackage{amsmath}
\usepackage{amssymb}
\usepackage{mathrsfs}

\setstcolor{yellow}

\begin{document}
\bibliographystyle{unsrt}
\preprint{APS/123-QED}
\title{Benchmarking the simplest slave-particle theory with Hubbard dimer}

\author{Wei-Wei Yang$^{1}$,Yin Zhong$^{1}$}
\email{zhongy@lzu.edu.cn}
\affiliation{%
 $^1$Center for Interdisciplinary Studies and Key Laboratory for Magnetism and Magnetic Materials of the MoE, Lanzhou University, Lanzhou 730000, People Republic of China} %
\author{Hong-Gang Luo$^{1,2}$}
\email{hgluo@lzu.edu.cn}
\affiliation{%
 $^1$Center for Interdisciplinary Studies and Key Laboratory for Magnetism and Magnetic Materials of the MoE, Lanzhou University, Lanzhou 730000, People Republic of China \\
$^2$Beijing Computational Science Research Center, Beijing 100084, China}%





\date{\today}
\setstcolor{yellow}
\begin{abstract}

Slave-particle method is a powerful tool to tackle the correlation effect in quantum many-body physics. Although it has been successfully used to comprehend various intriguing problems, such as Mott metal-insulator transition and Kondo effect, there is still no convincing theory so far on the availability and limitation of this method. The abuse of slave-particle method may lead to wrong physics. As the simplest slave-particle method, $\mathbb{Z}_2$ slave spin, which is widely applied to many strongly correlated problems, is highly accessible and researchable. In this work, we will uncover the nature of $\mathbb{Z}_2$ slave-spin method by studying a two-site Hubbard model. After exploring some properties of this toy model, we make a comparative analysis of the results obtained by three methods: (i) slave-spin method on mean-field level, (ii) slave-spin method with gauge constraint and (iii) the exact solution as a benchmark. We find that, protected by particle-hole symmetry, the slave-spin mean-field method can recover the static properties of ground state exactly at half filling. Furthermore, in the parameter space where both $U$ and $T$ are small enough, slave-spin mean-field method is also reliable in calculating dynamic and thermal dynamic properties. However, when $U$ or $T$ is considerably large, the mean-field approximation gives ill-defined behavior, which results from the unphysical states in enlarged Hilbert space. These findings lead to our conclusion that the accuracy of slave particle can be guaranteed if we can
exclude all unphysical states by enforcing gauge constraints. 
Our work demonstrates the promising prospect of slave-particle method in studying complex strongly correlated models with specific symmetry or in certain parameter space.    

\end{abstract}

\pacs{Valid PACS appear here}
\maketitle


\section{\label{sec1:level1}INTRODUCTION}
Correlation is a crucial ingredient responsible for many exotic phenomena,
 such as Mott insulator-metal transition \cite{1}, heavy fermions \cite{RevModPhys.79.1015}, unconventional superconductivity \cite{RevModPhys.63.239} and ultra-cold atoms in optical lattice \cite{RevModPhys.80.885}. However, these phenomena are difficult to study by traditional methods that base on independent particles approximation. Now we know that this challenge results from the entanglement of multiple degrees of freedom such as spin, charge and orbit that can not be captured by independent-particle representation.

 Fortunately, the invention of slave-particle concept enlightens physicists as to a novel method to systematically comprehend strong correlation physics. The main idea in such method is to enlarge the Hilbert space with auxiliary degrees of freedom so that the charge and spin degrees of freedom could be disentangled and thus many above-mentioned phenomena could be studied at a mean-field level. In the past several decades, slave-particle method was developed into several different forms, such as slave boson \cite{0305-4608-6-7-018,PhysRevLett.57.1362,PhysRevB.29.3035}, slave rotor \cite{PhysRevB.70.035114,PhysRevB.66.165111,PhysRevB.68.245311} and slave spin \cite{PhysRevB.72.205124,PhysRevB.81.155118}. These different slave-particle representations have their own advantages and are pioneered in tackling various strongly correlated electron problems \cite{0305-4608-6-7-018,PhysRevB.29.3035,PhysRevLett.57.1362,PhysRevB.72.205124,PhysRevLett.102.065301,PhysRevB.81.155118,PhysRevB.84.235115,0305-4608-6-7-018,PhysRevB.29.3035,PhysRevB.70.035114,PhysRevB.72.205124,PhysRevLett.102.065301,PhysRevB.81.155118,PhysRevB.83.165105,PhysRevB.81.035106,PhysRevB.84.235115,PhysRevB.86.085104,PhysRevB.96.115114,PhysRevB.70.035114,PhysRevB.70.035114,PhysRevB.92.235117,PhysRevLett.85.178}, e.g., Anderson and Kondo model \cite{0305-4608-6-7-018,PhysRevB.29.3035,PhysRevB.85.073106}, Mott localization \cite{PhysRevB.72.205124,PhysRevLett.102.065301,PhysRevB.81.155118,PhysRevB.70.035114}, quantum spin Hall effect \cite{PhysRevLett.108.046401,PhysRevLett.103.196803} and high temperature superconductivity \cite{RevModPhys.78.17}. 

In spite of these successful applications, the accuracy of these methods is uncontrollable due to the mean-field approximation or Gaussian fluctuation, and their reliability has not been studied quantitatively yet. As far as we know, there is neither systematic scheme nor detailed analyses to investigate the accuracy and limitation of slave-particle method. We believe that these are important for any unconventional method.
  Therefore, we will try to fill in this gap by studying a two-site Hubbard model with a specific slave-particle approach. In this paper, we consider the simplest $\mathbb{Z}_2$ slave-spin method, because it is easy to study due to the finite Hilbert space of slave spin. Its extension to other slave-particle method is straightforward. This method has predicted some new physics, such as orthogonal metals \cite{PhysRevB.86.045128,PhysRevB.86.115113} and fractionalized Chern insulator \cite{PhysRevB.88.045109}, and the former one has been confirmed in a recent quantum Monte carlo simulation \cite{PhysRevLett.121.086601}.
 We aim at making an explicit evaluation of $\mathbb{Z}_2$ slave-spin method by comparatively analyzing aspects of properties in this two-site Hubbard model, which is exactly soluble so that the benchmark is ready-made and we can evaluate slave spin fairly. 


 Here, we have two important conclusions:

(i) Some specific symmetries, e.g., particle-hole symmetry, can play important roles in reproducing exact behavior in slave-spin mean-field approach.

(ii) The accuracy of slave particle is ensured if only we can exclude all unphysical states by enforcing gauge constraints.

We demonstrate the reliability of the slave-spin method both on mean-field level and with gauge constraints at zero and finite temperature. It turns out that slave-spin mean-field method performs perfectly well in the static properties of the ground state, whereas its accuracy does not sustain at finite temperature. Since unphysical states are mixed in the eigenstates in this enlarged slave-spin representation, mean-field approximation cannot access dynamical or thermal dynamic behavior accurately. Fortunately, this ill-defined behavior can be fixed with enforcing gauge constraints exactly, i.e., excluding the unphysical states completely. Besides, the result of ground state also shows that slave-spin mean-field method is effective in both a small $U$ regime and a large $U$ regime, but it fails in the regime where $U$ is comparable to hopping integral $t$. For the latter case, the result either diverges or deviates far from the exact solution. By taking this $\mathbb{Z}_2$ slave-spin method as an example, we hope to uncover the nature of slave-particle theory and give some suggestions on the application and regime where the slave-particle method can work effectively and accurately.

\par The outline of this paper is as follows: at first we present the general formulation of $\mathbb{Z}_2$ slave-spin method in Hubbard model in Sec.\ref{sec2:level1}. In Sec.\ref{sec3:level1}, we introduce three theoretical approaches involved in the next context. We describe in Sec.\ref{sec3:level1A} the exact diagonalization method, in Sec.\ref{sec3:level1B} the $\mathbb{Z}_2$ slave-spin mean-field method, and in Sec.\ref{sec3:level1C} the slave-spin method with exact gauge constraints. In Sec.\ref{sec4:level1}, we demonstrate and analyze our results with above-mentioned three methods for different context: in Sec.\ref{sec4:level1A} and Sec.\ref{sec4:level1B}, the zero temperature situation, whereas in Sec.\ref{sec4:level1C} and Sec.\ref{sec4:level1D} the finite temperature situation. Within this section, the static, dynamical and thermal dynamic properties are discussed in various regime of $U$. Finally, we conclude in Sec.\ref{sec5:level1A}.

\section{\label{sec2:level1}Introduction to $\mathbb{Z}_2$ slave-spin method}

As a starting point, we introduce the main idea of $\mathbb{Z}_2$ slave spin in single-orbital Hubbard model \cite{0953-8984-27-39-393001,PhysRevA.88.062512,doi:10.1142/2762,2010odhm.book.....E,book3,book4}. The Hamiltonian can be written as follow:
\begin{eqnarray}
H&=&-\sum_{ij\sigma}t_{ij}c_{i\sigma}^{\dagger}c_{j\sigma}+U\sum_{i}n_{i\uparrow}n_{i\downarrow}-\mu\sum_{i\sigma}c_{i\sigma}^{\dagger}c_{i\sigma}.
\label{equ1}
\end{eqnarray}
The $c_{i\sigma}$ is electron annihilation operator on the site $i$ with spin $\sigma$; $n_{i\sigma}$ is the occupancy number of electrons which satisfies $n_{i\uparrow(\downarrow)}=c_{i\uparrow(\downarrow)}^{\dagger}c_{i\uparrow(\downarrow)}$; $t_{ij}$ is the hoping integral between $i$ site and $j$ site; $\mu$ is chemical potential. In the strongly correlated regime, the approximate methods based on independent particles are invalid. Therefore, we introduce the slave-spin method to solve this dilemma. In the slave-spin representation, the physical electron operator is written as \cite{PhysRevB.81.035106,PhysRevB.81.155118}:
\begin{eqnarray}
c_{i\sigma}&=&f_{i\sigma}\tau_x,
\label{equ2}
\end{eqnarray}
where $\tau_i^a (a=x, y, z)$ is the Pauli matrices which describe the auxiliary Ising spin, $f_{i\sigma}$ is the annihilation operator of the auxiliary fermion on site $i$ with spin $\sigma$. Note that the auxiliary fermions share the same anti-commutation relations with the physical fermions:

\begin{eqnarray}
\left\{f_{i\sigma},f_{j\sigma^{'}}\right\}=0,\left\{f_{i\sigma}^{\dagger},f_{j\sigma^{'}}^{\dagger}\right\}=0,\left\{f_{i\sigma},f_{j\sigma^{'}}^{\dagger}\right\}=\delta_{ij}\delta_{\sigma\sigma^{'}}.
\label{equ3}
\end{eqnarray}

 \begin{figure}
\centering
\includegraphics[width=3.5in]{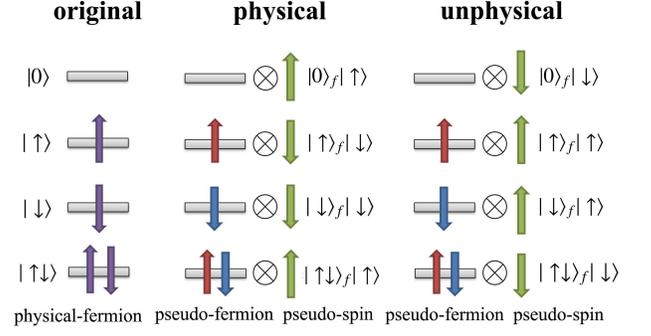}
\caption{The schematic illustration of $\mathbb{Z}_2$ slave-spin representation. The physical state of one site (left column) can be represented by pseudospin and pseudofermion. In the enlarged Hilbert space, half of states are physical (middle column) and another half are unphysical (right column). It is twice as large as the original one.}
\label{FIG1}
\end{figure}

As shown in Fig.\ref{FIG1}, with the discrete $\mathbb{Z}_2$ gauge symmetry, the Hilbert space is twice as large as the original one, i.e., in the $|fermions\rangle\otimes|Ising\rangle$ representation, only half of states are physical:
\begin{eqnarray}
|0\rangle=&\left|0\right\rangle_f|\uparrow\rangle,\nonumber\\
|\uparrow\rangle=&|\uparrow\rangle_f|\downarrow\rangle,\nonumber\\
|\downarrow\rangle=&|\downarrow\rangle_f|\downarrow\rangle,\nonumber\\
|\uparrow\downarrow\rangle=& |\uparrow\downarrow\rangle_f|\uparrow\rangle,
\label{equ3.1}
\end{eqnarray}
while another four states are unphysical:
\begin{eqnarray}
|0\rangle=&|0\rangle_f|\downarrow\rangle,\nonumber\\
|\uparrow\rangle=&|\uparrow\rangle_f|\uparrow\rangle,\nonumber\\
|\downarrow\rangle=&|\downarrow\rangle_f|\uparrow\rangle,\nonumber\\
|\uparrow\downarrow\rangle=&|\uparrow\downarrow\rangle_f|\downarrow\rangle.
\label{equ3.2}
\end{eqnarray}
Here $\tau_z|\uparrow(\downarrow)\rangle=\pm|\uparrow(\downarrow)\rangle$ in the pseudo-spin representation. To select physical states from unphysical ones, we have to apply some constraints at every site
\begin{eqnarray}
\tau_i^z=2\left(n_i^f-1\right)^2-1,
\label{equ4}
\end{eqnarray}
which can also be written as

\begin{eqnarray}
n_{i\uparrow}^fn_{i\downarrow}^f=\frac{1}{2}\left(\frac{\tau_i^z+1}{2}-1+n_i^f\right).
\label{equ4.1}
\end{eqnarray}
Here this non-linear constraint turns out to be a crucial ingredient to tackle with the cumbersome interaction term $n_{i\uparrow}^fn_{i\downarrow}^f$.

In this slave-spin formulation, we can write partition function as:
\begin{eqnarray}
\begin{split}
Z=\mathrm{Tr}(e^{-\beta{H}})=\mathrm{Tr}\left[P_Ie^{-\beta{H}_{slave-spin}}\right],
\label{6}
\end{split}
\end{eqnarray}
where $P_I$ is the projection operator onto the physical Hilbert space

\begin{eqnarray}
\begin{split}
P_I=\Pi_iP_I^i=\Pi_i\left(1-Q_i^2\right),
\label{7}
\end{split}
\end{eqnarray}

with
\begin{eqnarray}
\begin{split}
Q_i=\frac{\tau_i^z+1}{2}-\left(n_i^f-1\right)^2,
\label{8}
\end{split}
\end{eqnarray}
while only the states satisfying $Q_i=0$ are corresponding to physical states.


With $\mathbb{Z}_2$ slave-spin formulation [Eq.(\ref{equ2})] and constraint condition [Eq.(\ref{equ4})], we can translate Eq.(\ref{equ1}) into the slave-spin formulation
\begin{eqnarray}
\begin{split}
H_{slave-spin}=&-\sum_{ij\sigma}t_{ij}\tau_i^x\tau_j^{x}f_{i\sigma}^{\dagger}f_{j\sigma}
                +U\sum_{i}\frac{\tau_i^z-1}{4}\\
                &+\left(\frac{U}{2}-\mu\right)\sum_{i\sigma}f_{i\sigma}^{\dagger}f_{i\sigma}.
\label{5}
\end{split}
\end{eqnarray}
\par Now the complex strong correlation effect of this system is merely reflected in the dependence between the hopping integral $t$ of $f$ fermions and the coupling strength of nearest-neighbor pseudo-spins. Here, this transformed formalism is exact, but it looks cumbersome due to complicated hoping term. Generally, there are three methods to proceed:

(i) Upon most occasions, we can perform some mean-field approximations, with which the original Hamiltonian could be divided into three independent parts and then solved by iteration method.

(ii) For some small systems, we can make exact diagonalization and directly access to all properties.

(iii) For a considerable large system, quantum Mont Carlo simulation is widely used to obtain credible result.

For some extremely simple cases, such as the two-site model, it is also possible to apply slave spin with exact gauge constraints. In the next section, we set the second scheme to be the benchmark to justify the validity of slave spin at mean-field level and slave spin with gauge constraints.



\section{\label{sec3:level1}describe two-site Hubbard model with three approaches}

To further simplify the two-site model, we regard the system as a canonical ensemble, i.e., the chemistry potential should satisfy the constraint $\mu=0$. Now we start with the Hamiltonian

\begin{eqnarray}
H_{2-site}&&=-\sum_{\sigma}t\left(c_{1\sigma}^{\dagger}c_{2\sigma}+c_{2\sigma}^{\dagger}c_{1\sigma}\right)+U\left(n_{1\uparrow}n_{1\downarrow}+n_{2\uparrow}n_{2\downarrow}\right),\nonumber\\
&&\label{equ9}
\end{eqnarray}
where $n_{1(2)}=\sum_{\sigma}c_{1(2)\sigma}^{\dagger}c_{1(2)\sigma}$ is the electronic density on the site 1(2). We set hopping integral as energy scale and take $t=1$ throughout this paper. As the simplest possible example, this two-site Hubbard model has analytical solution by exact diagonalization. In the next subsection, we will investigate this system with exact solution, slave-spin method on mean-field level, and slave-spin method with exact gauge constraints, with which we could have a reasonable assessment of slave-spin method.


\subsection{\label{sec3:level1A}the analytical solution}
The analytical solution is directly derived from exact diagonalization without special skill. Hence, in this subsection we will neglect the simple non-half-filled situation but focus on the half-filled case, in which slave-spin mean-field approach surprisingly reproduces the behavior of exact solution. We hope to uncover the nature of this coincidence by discussing all methods at different cases in detail.
Now the starting point of our discussion of the canonical ensemble is this half-filling property, which means there should be two electrons in total. Thus we can directly list the entire basis vectors in the occupation number representation
\begin{eqnarray}
|\uparrow\uparrow\rangle&=c_{1\uparrow}^{\dagger}c_{2\uparrow}^{\dagger}|0\rangle\nonumber\\
|\downarrow\downarrow\rangle&=c_{1\downarrow}^{\dagger}c_{2\downarrow}^{\dagger}|0\rangle\nonumber\\
|\uparrow\downarrow\rangle&=c_{1\uparrow}^{\dagger}c_{2\downarrow}^{\dagger}|0\rangle\nonumber\\
|\downarrow\uparrow\rangle&=c_{1\downarrow}^{\dagger}c_{2\uparrow}^{\dagger}|0\rangle\nonumber\\
|0d\rangle&=c_{2\uparrow}^{\dagger}c_{2\downarrow}^{\dagger}|0\rangle\nonumber\\
|d0\rangle&=c_{1\uparrow}^{\dagger}c_{1\downarrow}^{\dagger}|0\rangle.
\label{equ9.1}
\end{eqnarray}
Because the states $|\uparrow\uparrow\rangle$, $|\downarrow\downarrow\rangle$ are the eigenstates of the Hamiltonian, i.e.,
\begin{eqnarray}
H|\uparrow\uparrow\rangle=&E|\uparrow\uparrow\rangle\nonumber\\
H|\downarrow\downarrow\rangle=&E|\downarrow\downarrow\rangle,
\label{equ9.1}
\end{eqnarray}
where $E=0$, we can neglect them in the subsequent analysis. Thus, the other four states constitute a 4-dimensional Hilbert space and the basis vector is

\begin{eqnarray}
|\psi\rangle=\left(
\begin{array}{c}
|\uparrow\downarrow\rangle\\
|\downarrow\uparrow\rangle\\
|0d\rangle\\
|d0\rangle
\end{array}
\right),
\label{equ10}
\end{eqnarray}
which is straightway to derive the Hamiltonian matrix in this occupation number representation
\begin{eqnarray}
\begin{split}
H_{2-site}=
\left(
\begin{array}{cccc}
0& 0& -t& -t\\
0& 0& t& t\\
-t& t& U& 0\\
-t& t& 0& U
\end{array}
\right).
\label{equ11}
\end{split}
\end{eqnarray}
By diagonalizing this Hamiltonian, we get four eigen-energies
\begin{eqnarray}
E&=&\frac{1}{2}\left(U-\sqrt{U^2+16t^2}\right),\nonumber\\
E&=&0,\nonumber\\
E&=&U,\nonumber\\
E&=&\frac{1}{2}\left(U+\sqrt{U^2+16t^2}\right).
\label{12}
\end{eqnarray}
As shown in Eq.(\ref{12}), the minimal energy solution, i.e., the ground-state energy, is $E_g=\frac{1}{2}\left(U-\sqrt{U^2+16t^2}\right)$. Thus we get the analytical solution for this 2-site Hubbard model.
In principle, with the help of Hellmann-Feynman theorem, now we can deduce most of physical quantities in ground state. For example, the double occupancy $d=\sum_{j\sigma}\langle c_{j\sigma}^{\dagger}c_{j\sigma}\rangle$ and bond order $\chi=\sum_{\sigma}\langle c_{1\sigma}^{\dagger}c_{2\sigma}+c_{2\sigma}^{\dagger}c_{1\sigma}\rangle$ of physical electrons can be derived directly from the analytical expression
\begin{eqnarray}
\begin{split}
\langle d\rangle_g&=\left\langle\frac{\partial H}{\partial U}\right\rangle_g=\frac{\partial E_g}{\partial U}=\frac{1}{2}\left(1-\frac{U}{\sqrt{U^2+16t^2}}\right),\\
\langle \chi\rangle_g&=-\left\langle\frac{\partial H}{\partial t}\right\rangle_g=-\frac{\partial E_g}{\partial t}=\frac{8}{\sqrt{U^2+16t^2}}.
\label{equ13}
\end{split}
\end{eqnarray}
\par When $U/t\rightarrow0$, the $E=0$ state and $E=U$ state are degenerate with each other and the ground-state energy $E_g\simeq-2t$, while the double occupancy number is $\frac{1}{2}$. It turns out that at a small $U$ limit, both two electrons are behaved as independent particles and the bond order of electrons is at its maximum value.

\par When $U/t\rightarrow\infty$, the $E=\frac{1}{2}\left(U-\sqrt{U^2+16t^2}\right)$ state and $E=0$ state are degenerate, while the $E=U$ state and $E=\frac{1}{2}\left(U+\sqrt{U^2+16t^2}\right)$ state are degenerate with each other. In this strong coupling limit, both the double occupancy $d$ and bond order $\chi$ tend to be zero. It is reasonable that when $U$ is increasing, the electrons prefer to occupy different sites to reduce the free energy of the system.

\subsection{\label{sec3:level1B}the $\mathbb{Z}_2$ slave-spin mean field}

In this subsection, we will explore the 2-site Hubbard model with the $\mathbb{Z}_2$ slave-spin mean-field method. With Eq.(\ref{equ2}) and Eq.(\ref{equ4.1}), the 2-site Hubbard Hamiltonian can be transformed into the slave-spin representation
\begin{eqnarray}
\begin{split}
H_{2-site}=&-t\tau_1^x\tau_2^x\sum_{\sigma}\left(f_{1\sigma}^{\dagger}f_{2\sigma}+f_{2\sigma}^{\dagger}f_{1\sigma}\right)\\
&+\frac{U}{2}\sum_{\sigma}\left(f_{1\sigma}^{\dagger}f_{1\sigma}+f_{2\sigma}^{\dagger}f_{2\sigma}\right)\\
&+\frac{U}{4}\left(\tau_1^z+\tau_2^z\right)-\frac{U}{2}.\\
\label{equ13.1}
\end{split}
\end{eqnarray}
To tackle with the four operator $\tau_1^x\tau_2^xf_{1\sigma}^{\dagger}f_{2\sigma}$ in this new Hamiltonian, next we take the mean-field approximation
\begin{eqnarray}
\tau_1^x\tau_2^xf_{1\sigma}^{\dagger}f_{2\sigma}=&\left<\tau_1^x\tau_2^x\right>f_{1\sigma}^{\dagger}f_{2\sigma}+\tau_1^x\tau_2^x\left<f_{1\sigma}^{\dagger}f_{2\sigma}\right>\nonumber\\
&-\left<\tau_1^x\tau_2^x\right>\left<f_{1\sigma}^{\dagger}f_{2\sigma}\right>.
\label{equ14}
\end{eqnarray}
Thus the auxiliary Ising spin operator and the auxiliary fermion operator are decoupled with each other.
Generally, it is hard to deal with the constraint precisely on every site, so we can also take constraint condition into account at mean-field level, i.e., adding some Lagrangian multipliers in original Hamiltonian
\begin{eqnarray}
\begin{split}
\sum_i\lambda_i\left(\frac{\tau_i^z+1}{2}-(n_i^f-1)^2\right).
\label{equ13}
\end{split}
\end{eqnarray}
Due to the translation invariance of the Hamiltonian, in this 2-site situation we simply assume $\lambda_{i(i=1,2)}=\lambda$. Now the Hamiltonian takes the form $H_{2-site}=H_f+H_{Ising}+E_0$, where
\begin{eqnarray}
\begin{split}
H_f=&-tI_{12}\sum_{\sigma}\left(f_{1\sigma}^{\dagger}f_{2\sigma}+f_{2\sigma}^{\dagger}f_{1\sigma}\right)+\frac{U}{2}\sum_{\sigma}\left(f_{1\sigma}^{\dagger}f_{1\sigma}+f_{2\sigma}^{\dagger}f_{2\sigma}\right)\\
&-\lambda\left((n_1^f-1)^2+(n_2^f-1)^2\right),\\
H_{Ising}&=-t\chi_{12}\tau_1^x\tau_2^x+\left(\frac{U}{4}+\frac{\lambda}{2}\right)\left(\tau_1^z+\tau_2^z\right),\\
E_0=&-\frac{U}{2}+\lambda+tI_{12}\chi_{12}.
\label{equ15}
\end{split}
\end{eqnarray}
 Here the mean-field parameters $I_{12}=\langle\tau_1^x\tau_2^x\rangle$ and $\chi_{12}=\sum_{\sigma}\langle f_{1\sigma}^{\dagger}f_{2\sigma}+f_{2\sigma}^{\dagger}f_{1\sigma}\rangle$. Note that at half filling $\lambda$ should be zero to respect the particle-hole symmetry, where the energy spectrum is symmetric only if $\lambda=0$. Under this assumption the Hamiltonian can be written as
\begin{eqnarray}
\begin{split}
&H_f=-tI_{12}\sum_{\sigma}\left(f_{1\sigma}^{\dagger}f_{2\sigma}+f_{2\sigma}^{\dagger}f_{1\sigma}\right)+\frac{U}{2}\sum_{\sigma}\left(f_{1\sigma}^{\dagger}f_{1\sigma}+f_{2\sigma}^{\dagger}f_{2\sigma}\right)\\
&H_{Ising}=-t\chi_{12}\tau_1^x\tau_2^x+\frac{U}{4}\left(\tau_1^z+\tau_2^z\right),\\
&E_0=-\frac{U}{2}+tI_{12}\chi_{12}.
\label{equ15}
\end{split}
\end{eqnarray}
In the physical ground state, the pseudo-fermions term $H_f$ should be self-consistent with the pseudo-spins $H_{Ising}$. In order to achieve this self-consistency, at first we have to set an initial value for $I_{12}$. With this $I_{12}$ we can derive the Hamiltonian of the $f$ fermions at a specific $U$. Afterwards, we calculate the eigenstates $|\phi_f\rangle$ and then $\chi_{12}=\sum_{\sigma}\langle\phi_f|f_{1\sigma}^{\dagger}f_{2\sigma}+f_{2\sigma}^{\dagger}f_{1\sigma}|\phi_f\rangle$, with which we can deduce the Hamiltonian matrix and thus the eigenstates of pseudo spins $|\phi_{Ising}\rangle$. Next we get a new $I_{12}=\sum_{\sigma}\langle\phi_{Ising}|\tau_1^x\tau_2^x|\phi_{Ising}\rangle$. Now we update the initial $I_{12}$ with the new one calculated from the pseudo-spins system. By repeating these steps mentioned above, we can finally get a self-consistent result in required accuracy range. Generally, for this simple two-site half-filling model it takes less than 30 steps to reach the convergence.


Note that,
 the ground-state energy could be directly written as
 \begin{eqnarray}
\begin{split}
E_g=\left<H_f\right>+\left<H_{Ising}\right>+E_0=E_f+E_{Ising}+E_0.
\label{equ15.1}
\end{split}
\end{eqnarray}
From the constraint Eq.(\ref{equ4.1}) we can derive the double occupancy in slave-spin mean-field level
\begin{eqnarray}
\begin{split}
d&=\frac{\tau_1^z+\tau_2^z}{4}+\frac{n_1^f+n_2^f}{2}-\frac{1}{2},
\label{equ16}
\end{split}
\end{eqnarray}
and also the bond order of physical particles
\begin{eqnarray}
\begin{split}
\chi_e&=I_{12}\chi_{12}.
\label{equ17}
\end{split}
\end{eqnarray}

\subsection{\label{sec3:level1C}the $\mathbb{Z}_2$ slave-spin method with exact gauge constraints}
As mentioned above, this slave-spin representation is respect to $\mathbb{Z}_2$ gauge symmetry \cite{PhysRevD.19.3682,RevModPhys.51.659}. 
Thanks to the simplicity of both the gauge symmetry and the two-site model, in this subsection we could derive an exact solution in the enlarged Hilbert space without mean-field approximation.

Next we make exact diagonalization of Hamiltonian [Eq.(\ref{equ13.1})] in this four times enlarged representation and derive to 64 eigenstates
\begin{eqnarray}
\begin{split}
H_{two-site}|\psi_{mix}\rangle=E_{mix}|\psi_{mix}\rangle.
\label{equ17.5}
\end{split}
\end{eqnarray}
Here we label its eigenstates as $|\psi_{mix}\rangle$, because only one-fourth of them are physical states. To select these physical states from the unphysical ones, we have to construct gauge transformation with appropriate formulation whose eigenstates could be utilized to distinguish physical states. Considering the constraint Eq.(\ref{equ4}), we define a local gauge transformation at site $i$ as the operator $G(i)=\tau_i^z(-1)^{n_i}$.

Firstly, we will demonstrate the reasonability of this gauge transformation. For example in the first term of Eq.(\ref{equ13.1}),
\begin{eqnarray}
\begin{split}
\tau_i^z\tau_1^x\tau_2^x\left(\tau_i^z\right)^{-1}=-1,\\
\left(-1\right)^{n_i}\sum_{\sigma}\left(f_{1\sigma}^{\dagger}f_{2\sigma}+f_{2\sigma}^{\dagger}f_{1\sigma}\right)\left(-1\right)^{n_i}=-1.
\label{equ17.1}
\end{split}
\end{eqnarray}
The first equation is derived from the anti-commutation relations of pseudo-spins; the second demonstrates that hopping terms in fact change pairity of occupied number on every site. In a similar way, the other three terms are also gauge-invariants under this transformation, i.e.,
\begin{eqnarray}
\begin{split}
G(i)H_{two-site}\left(G(i)\right)^{-1}=H_{two-site}.
\label{equ17.2}
\end{split}
\end{eqnarray}
Thus we prove the commutative relation between Hamiltonian and operator G(i).

Next we discuss the idea of how to accurately exclude unphysical states. Notably, \begin{eqnarray}
\begin{split}
G(i)^2=1,
\label{equ17.3}
\end{split}
\end{eqnarray}
so that the eigenstate $|\phi_i>$ of operator G(i) satisfies
\begin{eqnarray}
G(i)|\phi_i\rangle=\pm|\phi_i\rangle,
\label{equ17.4}
\end{eqnarray}
where only $+1$ is corresponding to the physical states. Since the operator G(i) is commuted with Hamiltonian, they must have the same set of eigenstates $|\theta\rangle$. Here we find such a set and select only 16 physical states which satisfy $G(1)|\theta\rangle=|\theta\rangle,G(2)|\theta\rangle=|\theta\rangle$ to recover the behavior of this two-site model.

Although the Hamiltonian is gauge-invariant, the ground state has been changed when we enlarge initial Hilbert space. The physical ground state is still one of 64 eigenstates, but it no longer possesses the lowest energy. This will be shown in next section that the lowest energy is invariable with different $U$ if we simply describe the system with wave-function mixed with physical and unphysical states. Obviously, the unphysical states badly impact all aspects of behavior. If constraint conditions originally satisfied, mean-field approximation could eliminate the effect of unphysical states and access to some static properties effectively, such as the ground state. While it comes to the dynamical properties or finite temperature, the deviation is stubborn at mean-field level. To demonstrate the great influence caused by unphysical states, in next section we will show also the results of entire 64 states without gauge constraints.

\section{\label{sec4:level1}result and analysis}

\subsection{\label{sec4:level1A}the ground-state properties and analysis}
\begin{figure*}
\centering
\includegraphics[width=7in]{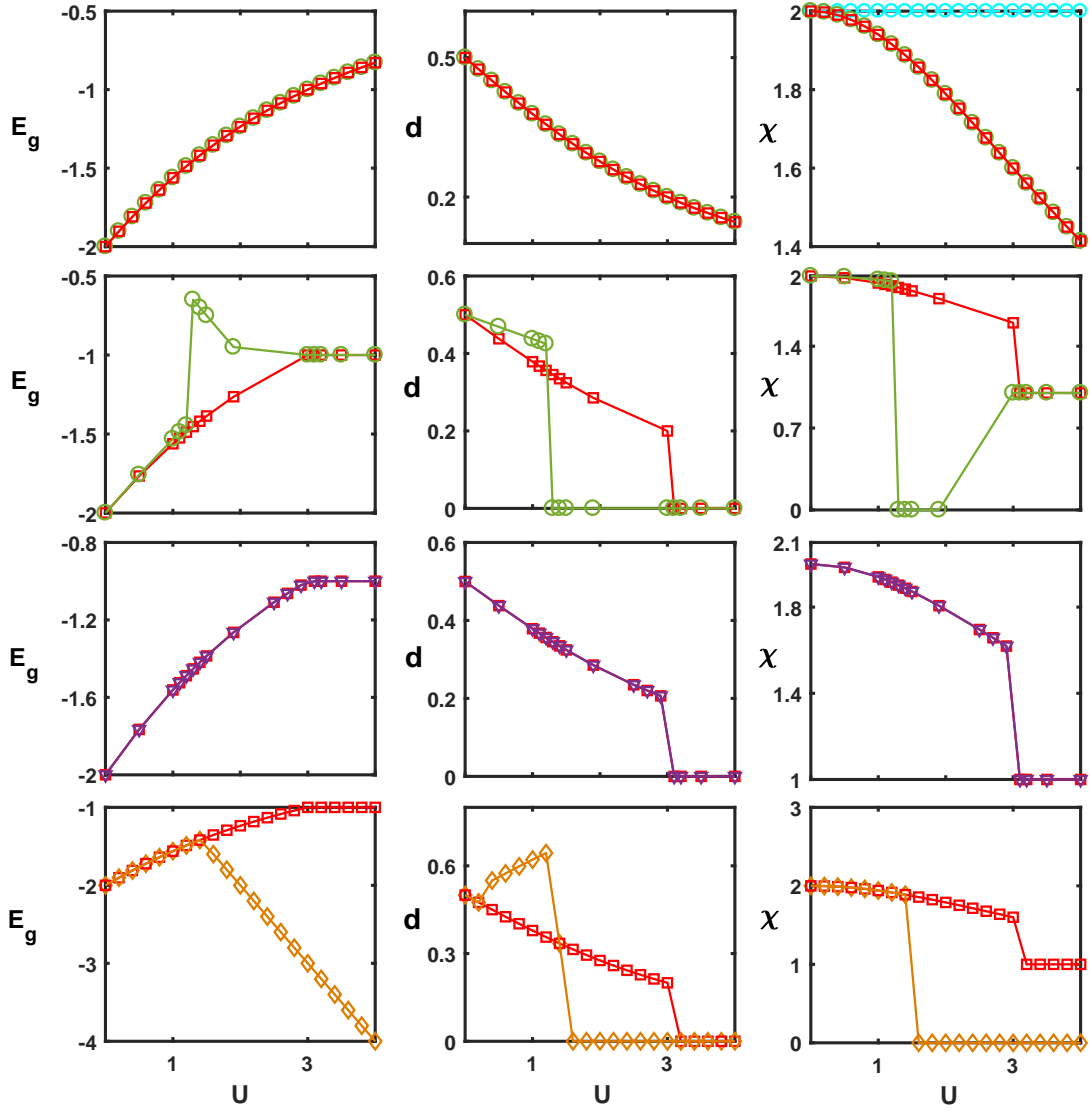}
\caption{The comparison of ground-state energy $E_g$ (left), double occupancy $d$ (middle) and bond order of physical fermions $\chi$ (right) of 2-site Hubbard model. The first row: Slave-spin mean-field (MF) method (green circle) and analytical solution (red square) in half-filling situation. Besides, the bond order of pseudo-fermions (cyan circle) is also demonstrated in the third row. The second row: Slave-spin MF method (green circle) and analytical solution (red square) with arbitrary electron number. The third row: Slave-spin method with gauge constraints (GC) (blue triangle) and analytical solution (red square) with arbitrary electron number. The fourth row: Slave-spin method without GC (yellow diamond) and analytical solution (red square) with arbitrary electron number.}
\label{FIG2}
\end{figure*}

 \begin{figure}
\centering
\includegraphics[width=3.5in]{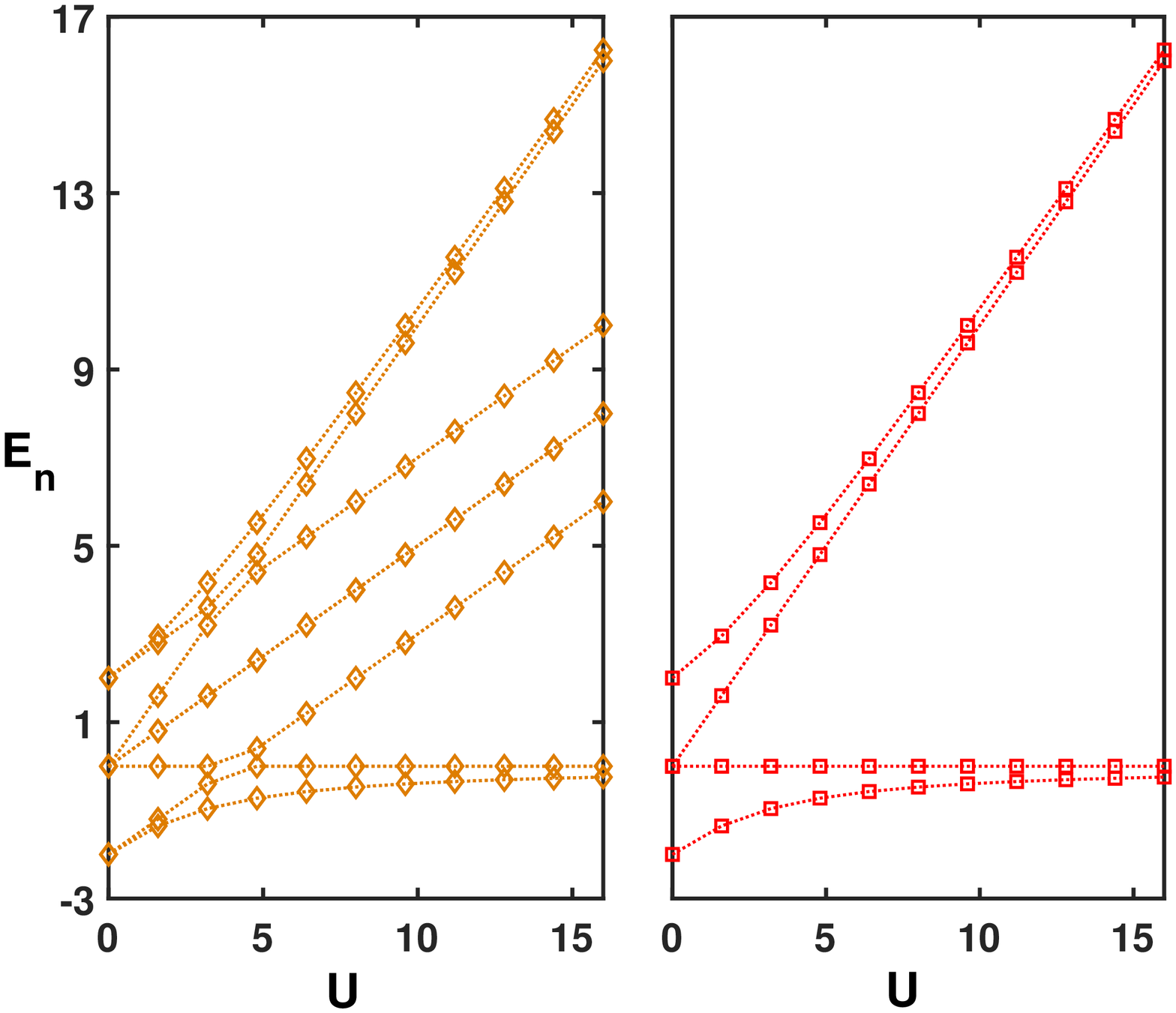}
\caption{At half filling, the comparison of energy spectrum distribution in slave-spin representation (yellow diamond) and original physical representation (red square).}
\label{FIG2.5}
\end{figure}

 \begin{figure}
\centering
\includegraphics[width=3.5in]{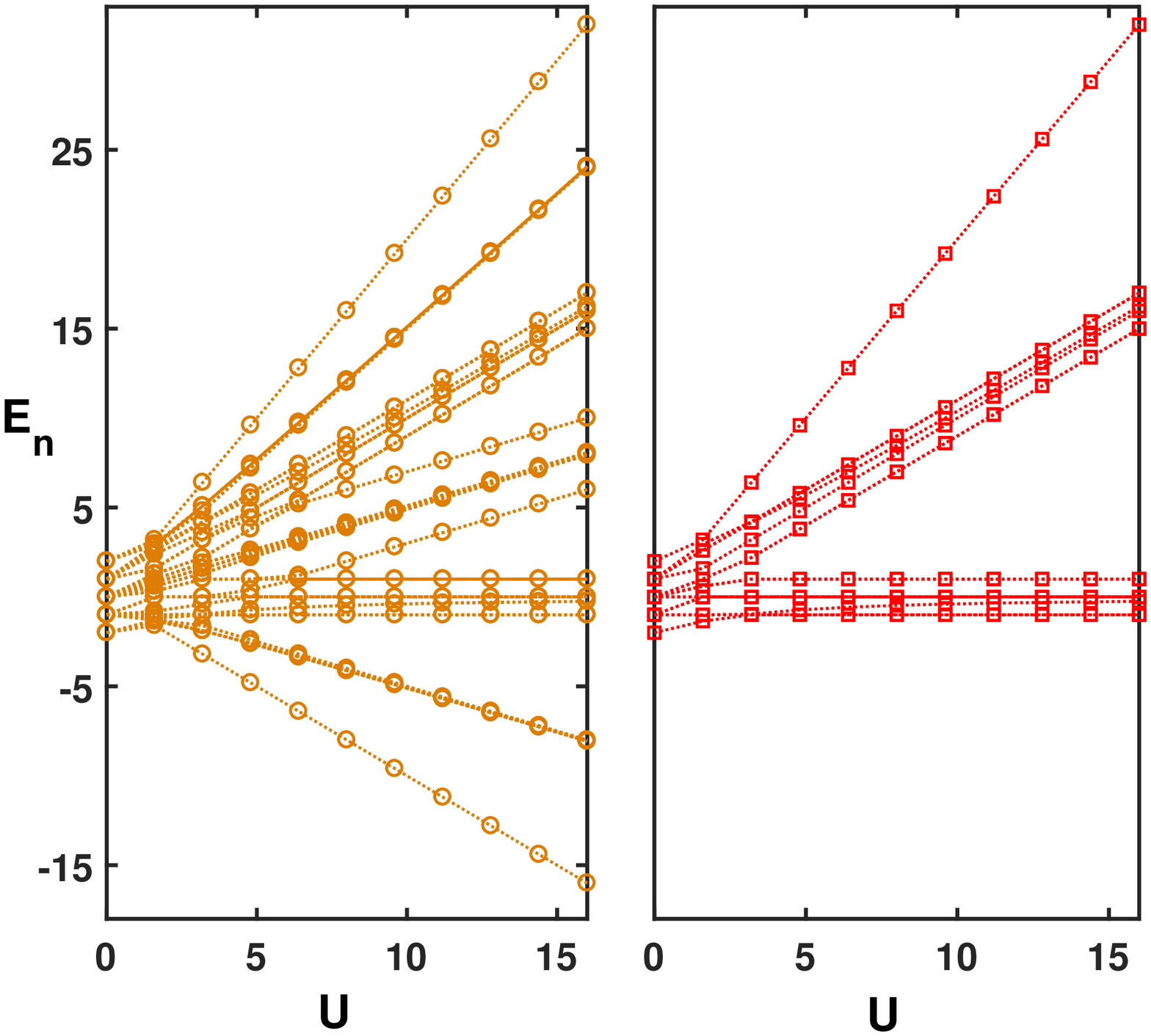}
\caption{At non-half filling, the comparison of energy spectrum distribution in slave-spin representation (yellow diamond) and original physical representation (red square).}
\label{FIG2.6}
\end{figure}

In this subsection, we demonstrate and discuss the performance of slave-spin approach by calculating typical static properties of ground state, such as energy, double occupancy and bond order of physical fermions. It turns out in some contexts slave spin can describe the 2-site model extremely accurate whereas in some other situations great deviation emerges. In the following, we will focus on revealing the kernel hinge which decides the efficiency of slave-spin method by contrasting different methods in different situations, half-filled or non-half-filled.

We first address the performance of slave spin on mean-field level in the half-filling situation. Figure \ref{FIG2} compares different slave-spin methods and analytical solution. We provide the dependence of static properties of ground state on $U$, in which slave-spin mean-field model predicts exactly the same results with analytical solution. The accuracy is not caused by accident but the particle-hole symmetry. As mentioned in the last section, Lagrange multipliers should be zero, i.e., $\lambda=0$, respecting to this special symmetry. It means that the form of Hamiltonian is fixed even if we have to consider the gauge constraint. As constraint conditions will not impact Hamiltonian directly, we can simply start from the Hamiltonian in the enlarged Hilbert space without any gauge constraints but access to an accurate behavior of ground state.

This is not the first show of particle-hole symmetry in slave-spin method. In both Anderson impurity model \cite{PhysRevB.81.155118,
PhysRevB.87.037101,PhysRevB.87.037102} and the infinite-dimensional half-filled Hubbard model \cite{PhysRevB.91.245130}, which can map with each other with dynamical mean-field theory \cite{RevModPhys.68.13}, their partition function can be calculated with corresponding slave-spin Hamiltonian without any constraint. Since the specific form of partition functions is irrelevant to the constraint conditions, so is the free energy. It turns out that slave-spin method is elegant and effective in these models which possesses particle-hole symmetry.

Since this fortune results from the special half-filling situation, in more general context slave spin cannot access to such perfect results with mean-field approximation. We demonstrate properties of two-site model with arbitrary electron number (see second row in Fig.\ref{FIG2}). Since in this situation constraint conditions could not be enforced exactly, deviation shows up at a small $U$ regime. Nevertheless, the slave-spin mean-field model has a more ill-defined behavior at moderate $1<U/t<3$ regime. Without particle-hole symmetry, mean-field approximation is not accurate anymore. Finally, figure \ref{FIG2} illustrates that slave-spin approach employing exact gauge constraints automatically corrects the ill-defined behavior at moderate $U$ regime.

Here we also provide the result from the enlarged slave-spin representation with physical and unphysical states mixed together (see the fourth row in Fig.\ref{FIG2}). Unfortunately, without constraints the behavior of slave spin is badly influenced by unphysical states and the result seems to be nonsense at most regime. Therefore, in next subsection we no longer demonstrate the result with this non-constraint slave-spin method.

As shown in Fig.\ref{FIG2.5} and Fig.\ref{FIG2.6}, we plot the energy spectrums of original physical representation and slave-spin representation at half filling and non-half filling respectively. It turns out that all the physical states are included in the four times enlarged Hilbert space. Notably, at small $U$ regime, the distributions of eigen-energy are almost the same in both two representations, whereas at large $U$ limit the fake sates deviated greatly from the physical ones. As a matter of fact, the feature of energy spectrums is accounting for the the gratified results at small $U$ regime and the ill-defined results at large $U$ regime of slave-spin method in calculating thermal dynamic properties (as shown in Sec.\ref{sec4:level1D}).

Obviously, in non-half-filling situation, at most large $U$ regime the ground state is not the physical state at all (see Fig.\ref{FIG2.6}). In general cases, we simply estimate the stability of system according to the level of energy, such as in the Variational Monte Carlo method. However, in some artificial representations, even the ground state may be fake states. The height of energy level should not be the only ruler to judge the ground state, where we should also ensure the inherent physical nature of the objects discussed. Conversely, guaranteed by particle-hole symmetry, at half filling the slave-spin method reproduces the exactly right ground state as analytical solution (see Fig.\ref{FIG2.5}). In both situations, the unphysical states which possess greatly deviated eigen-energies take up a much bigger proportion than physical ones. Most excited states are unphysical so that the finite-temperature framework is not so reliable without stringent constraints, where these unphysical states will impact observables heavily, especially at large $U$ regime.

As indicated above, the crucial ingredient to ensure the accuracy of slave-spin approach is excluding the unphysical states by exactly enforcing gauge constraints. Next we will analyze the processing of iteration in detail to elucidate how particle-hole symmetry plays such an important role in this slave-particle method. Due to the fact of half-filling, the total electron number should be an invariant to respect the particle-hole symmetry, i.e., $\sum_{i,\sigma}f_{i\sigma}^{\dagger}f_{i\sigma}=2$. Thus in the original Hamiltonian [Eq.(\ref{equ15})] pseudo-fermions term $H_f$ could now be rewritten in an equivalent form to give
\begin{eqnarray}
\begin{split}
&H_f=-tI_{12}\sum_{\sigma}\left(f_{1\sigma}^{\dagger}f_{2\sigma}+f_{2\sigma}^{\dagger}f_{1\sigma}\right),\\
&H_{Ising}=-t\chi_{12}\tau_1^x\tau_2^x+\frac{U}{4}\left(\tau_1^z+\tau_2^z\right),\\
&E_0=\frac{U}{2}+tI_{12}\chi_{12}.
\label{equ18}
\end{split}
\end{eqnarray}
  It turns out that no matter what initial value has been set for the $I_{12}$, the ground-state wave-function of the auxiliary fermions is unchangeable so that we get an invariant $\chi_{12}$. A fixed $\chi_{12}$ also leads to a fixed ground-state wave-function for the auxiliary Ising spin degree of freedom. Thus we directly get a self-consistent ground state, while the iteration process is not indeed necessary. As shown in Fig.\ref{FIG2}, now the bond order of quasi-particle $\chi_{12}$ is a constant irrelevant to the correlation strength $U$. As a result, the average bond order of physical electrons $\chi=I_{12}\chi_{12}$ is only dependent on the auxiliary spin degree of freedom. It turns out that a bond order only depending on a single degree of freedom, i.e., the thorough decoupling between different slave variables, is a crucial signal which suggests the slave-spin method may possess a precise solution of ground state.

As to the two-site model with arbitrary electron number, the iteration process is much more complex.
 Different from half-filling context, now the Hamiltonian of pseudo-fermions written as
\begin{eqnarray}
\begin{split}
&H_f=-tI_{12}\sum_{\sigma}\left(f_{1\sigma}^{\dagger}f_{2\sigma}+f_{2\sigma}^{\dagger}f_{1\sigma}\right)+\frac{U}{2}\sum_{\sigma}\left(f_{1\sigma}^{\dagger}f_{1\sigma}+f_{2\sigma}^{\dagger}f_{2\sigma}\right),\\
\\
\label{equ18.1}
\end{split}
\end{eqnarray}
which indicates that both $H_f$ and $H_{Ising}$ in Eq.(\ref{equ15}) are all dependent on correlation strength $U$. Therefore, sufficient iterations are inevitably needed to achieve a self-consistent ground state. Now the physical bond order is relying on both charge and spin degrees of freedom. The entanglement between pseudo-spins and pseudo-fermions naturally leads the slave-spin solution on mean-field level deviating from the analytical solution.


  %


 \subsection{\label{sec4:level1B}quantum fluctuations beyond mean field}

 \begin{figure}
\centering
\includegraphics[width=2.5in]{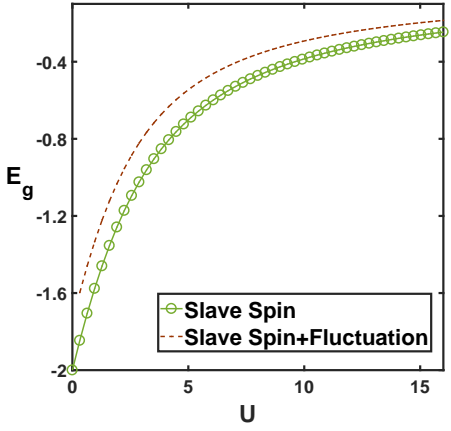}
\caption{The comparison of ground-state energy with slave-spin MF method (green circle) and the slave-spin method including the effect of fluctuation (brown dotted-line).}
\label{FIG3}
\end{figure}
In this subsection, we will additionally demonstrate some extremely intriguing properties of slave-spin approach in half-filling situation. So far, we merely apply this slave-spin method to investigate this two-site Hubbard model on a mean-field level. Even if we have made some approximations for convenience, the slave spin still performs exactly the same with analytical solution, which makes us wonder what if we further consider the effect of quantum fluctuations beyond mean-field and whether this system would be described more accurately.
\par In the last several subsections, after making transformation to the slave-spin representation, we generally tackle the four operators $\tau_1^x\tau_2^xf_1^{\dagger}f_2$ term with mean-field theory and neglect any quantum fluctuations. Now there is an alternative method to go beyond this local approximation with the perturbation theory, where we mainly consider the quantum fluctuations of the transverse Ising part after making mean-field approximation. In order to apply the perturbation theory, it is necessary to change the Hamiltonian into the form
\begin{eqnarray}
\begin{split}
H_{slave-spin}=H_0+H_{int},
\label{equ18.4}
\end{split}
\end{eqnarray}
where $H_0$ still takes the slave-spin formulation:
\begin{eqnarray}
\begin{split}
&H_f=-tI_{12}\sum_{\sigma}\left(f_{1\sigma}^{\dagger}f_{2\sigma}+f_{2\sigma}^{\dagger}f_{1\sigma}\right)+\frac{U}{2}\sum_{\sigma}\left(f_{1\sigma}^{\dagger}f_{1\sigma}+f_{2\sigma}^{\dagger}f_{2\sigma}\right)\\
&H_{Ising}=-t\left<\sum_{\sigma}\left(f_{1\sigma}^{\dagger}f_{2\sigma}+f_{2\sigma}^{\dagger}f_{1\sigma}\right)\right>\tau_1^x\tau_2^x\\
&E_0=\frac{U}{2}+t\chi_{12}I_{12},\\
\label{equ18.5}
\end{split}
\end{eqnarray}
and the perturbation term is written as:
\begin{eqnarray}
\begin{split}
H_{int}=&-t\left(\tau_1^x\tau_2^x-\left<\tau_1^x\tau_2^x\right>\right)\sum_{\sigma}\left(f_{1\sigma}^{\dagger}f_{2\sigma}+f_{2\sigma}^{\dagger}f_{1\sigma}\right).
\label{equ18.6}
\end{split}
\end{eqnarray}
To get the ground-state properties with correction up to second order, firstly we derive the ground wave function with first-order correction
\begin{eqnarray}
|\psi_g^{(1)}\rangle=\sum_{n\ne0}\frac{H_{n0}^{'}}{E_g-E_n}|\psi^{(0)}_n\rangle,
\label{equ18.7}
\end{eqnarray}
where the index $n$ stands for the seven excited states in the slave-spin representation, which are calculated from the non-perturbation Hamiltonian $H_0$; $|\psi^{(0)}_n\rangle$ is the $n-th$ excited eigenstate without correction. As we can see, the first-order correction $E^{(1)}_g=\langle\psi_g^{(0)}|H_{int}|\psi_g^{(0)}\rangle=0$. Next with first-order corrected wave-function, we could further calculate the second-order correction
\begin{eqnarray}
E^{(2)}_g=\langle\psi_g^{(0)}|H_{int}|\psi_g^{(1)}\rangle.
\label{equ18.4}
\end{eqnarray}
As shown in Fig.\ref{FIG3}, we plot the energy including quantum fluctuations beyond mean-field in varying $U$ and compare it with the original mean-field result. Intriguingly, this figure demonstrates that the quantum fluctuations beyond mean-field lead the energy deviating from the exact solution. The effect of fluctuations seems to be monotonously elevating the energy of the slave-spin mean-field solution.
 \par In this two-site Hubbard model, there are two reasons that make the mean-field solution with fluctuation correction even not so precise as the mean-field solution. First of all, we only include the first-order and second-order correction but neglect all higher-order ones. The first and second order corrections lift the ground-state energy, while these higher-order corrections sum up to a pure decline effect. As a result, if we consider the corrections of all orders, coincidently they conspire to make no difference, thus make the mean-field level solution as precise as the exact solution. What's more, we calculate the corrections in the enlarged Hilbert space, i.e., there are some unphysical states mixed with physical ones in the corrected ground wave function. The unphysical states make it hard to accurately calculate the excited-state wave-functions and also the effect of quantum fluctuations beyond MF. Although it has been proved that in some special situations we can access to the expected value of some physical properties without accurate constraint \cite{PhysRevB.87.037102,PhysRevB.87.037101}, the deviation somehow is unavoidable when considering fluctuations beyond mean-field or generic time evolution of observables.

 \subsection{\label{sec4:level1C}dynamics: Green function with time evolution}

In last section, we discuss both advantages and disadvantages of slave spin in describing ground state in various contexts, which are limited to static properties in zero temperature situations. In this section, we start from single-particle Green function, and focus on how different parameters impacting the accuracy of slave-spin method with time evolution. These parameters include temperature $T$, correlation strength $U$, and also electron number $n_f$.
The Green function generally reads as
\begin{eqnarray}
\begin{split}
G_{\sigma}\left(\emph{r}_i,\emph{r}_j;\tilde{t}\right)=-i\left<Tc_{j\sigma}\left(\tilde{t}\right)c_{i\sigma}^{\dagger}\left(0\right)\right>.
\label{equ18.1}
\end{split}
\end{eqnarray}
Here $T_t$ is the time-ordering operator and $\tilde{t}$ is the real time. If we use the slave-spin method and consider exact gauge constraints, Green function could be calculated with the eigenstates in the enlarged Hilbert space directly. However, to keep gauge constraints satisfied, it is necessary to get rid of unphysical states in wave function at every period. Otherwise, if we make mean-field approximation in this slave-spin representation, Eq.(\ref{equ18.1}) changes to
\begin{eqnarray}
\begin{split}
G_{\sigma}\left(\emph{r}_i,\emph{r}_j;\tilde{t}\right)\approx B_{ij}\left(\tilde{t}\right)G_{\sigma}^f\left(\emph{r}_i,\emph{r}_j;\tilde{t}\right).
\label{equ18.2}
\end{split}
\end{eqnarray}
Here we we have introduced some auxiliary quantities
\begin{eqnarray}
\begin{split}
B_{ij}\left(\tilde{t}\right)&=\langle TI_j^x\left(\tilde{t}\right)I_i^x(0)\rangle,\\
G_{\sigma}^f\left(\emph{r}_i,\emph{r}_j;\tilde{t}\right)&=-i\langle Tf_{j\sigma}\left(\tilde{t}\right)f_{i\sigma}^{\dagger}(0)\rangle,
\label{equ18.3}
\end{split}
\end{eqnarray}
where $B_{ij}\left(\tilde{t}\right)$ and $G_{\sigma}^f\left(\emph{r}_i,\emph{r}_j;\tilde{t}\right)$ could be calculated in sub-space separately.

\begin{figure}
\centering
\includegraphics[width=3.5in]{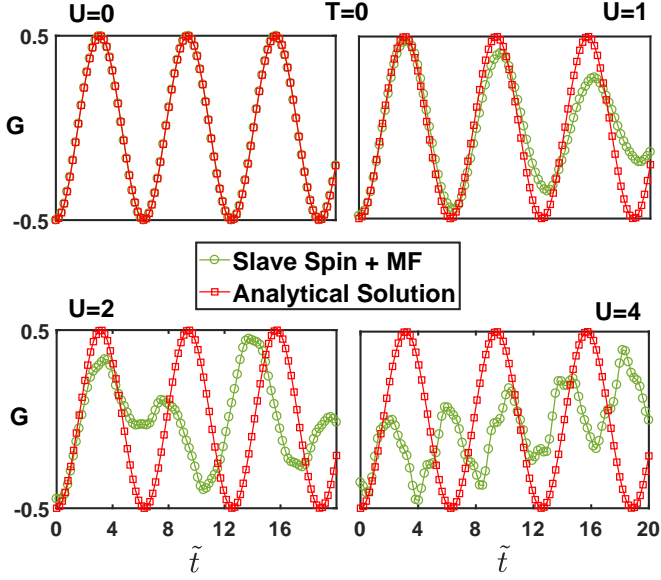}
\caption{The single-particle Green function $G(\tilde{t})$ with evolution of time in zero-temperature half-filling situation. Exact diagonalization (red square) and the $\mathbb{Z}_2$ slave-spin MF method (green circle). The results are calculated with different $U$.}
\label{FIG4}
\end{figure}

\begin{figure}
\centering
\includegraphics[width=3.5in]{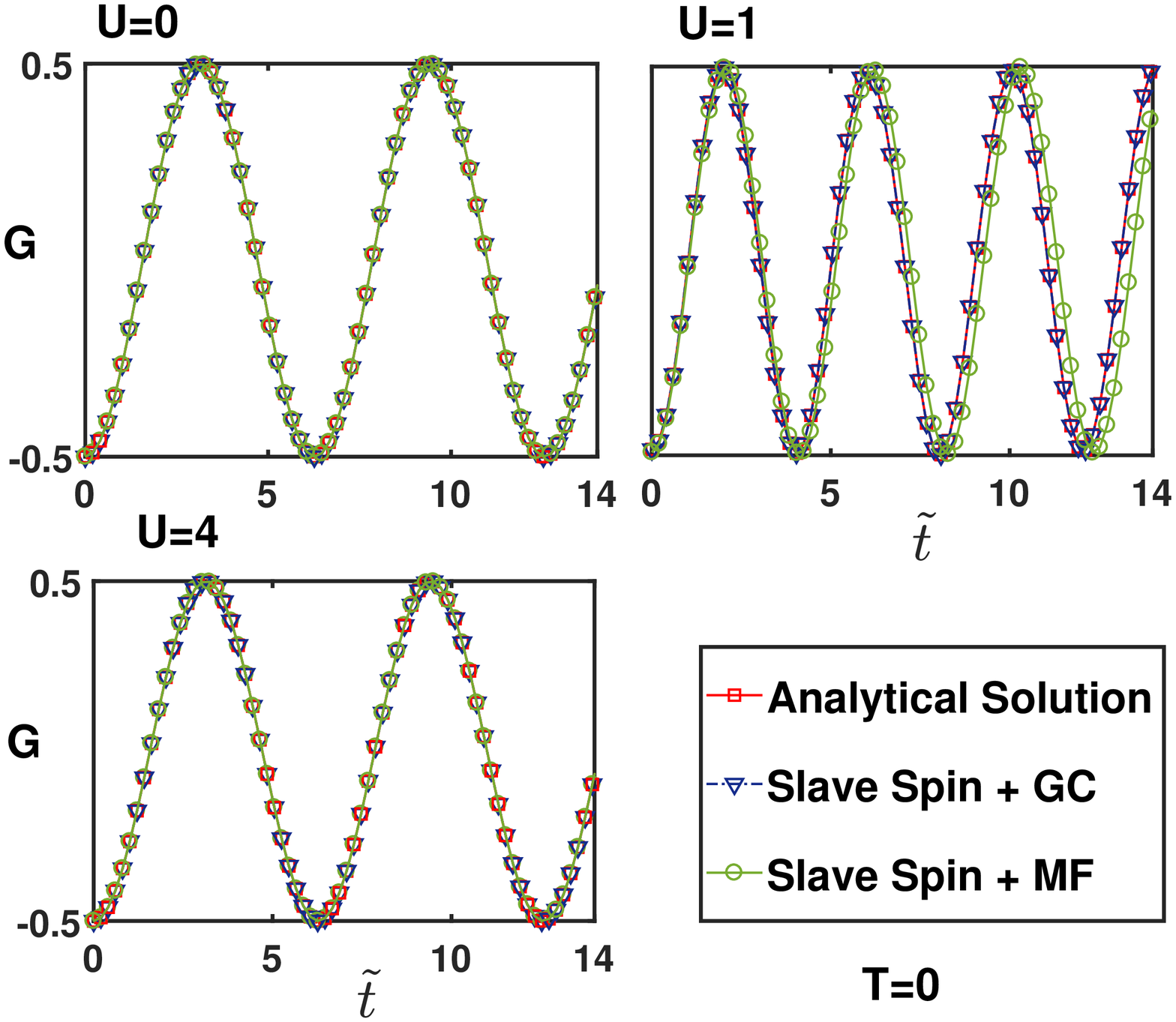}
\caption{The single-particle Green function $G(\tilde{t})$ with evolution of time in zero-temperature non-half-filled situation. Exact diagonalization (red square), the $\mathbb{Z}_2$ slave-spin MF method (green circle) and the slave-spin method with GC (blue triangle). The results are calculated with different $U$.}
\label{FIG5}
\end{figure}

\begin{figure}
\centering
\includegraphics[width=3.5in]{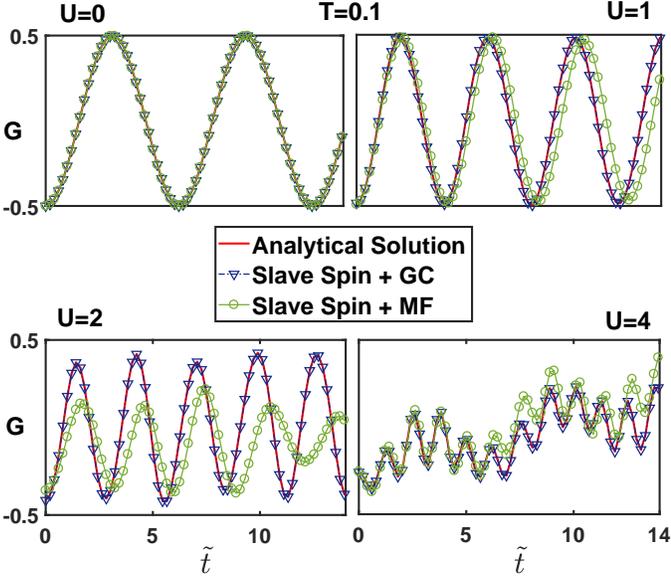}
\caption{The single-particle Green function $G(\tilde{t})$ with evolution of time in finite-temperature ($kT/t=0.1$) non-half-filled situation. Exact diagonalization (red line), the $\mathbb{Z}_2$ slave-spin MF method (green circle) and the slave-spin method with GC (blue triangle). The results are calculated with different $U$.}
\label{FIG6}
\end{figure}


Firstly we demonstrate the zero temperature situation, i.e., the dynamics of ground state, in half-filled (see in Fig.\ref{FIG4}) and non-half-filled (see in Fig.\ref{FIG5}) situation. Unfortunately, only if there is no interaction between electrons, the slave-spin mean-field model could exactly reproduce the standard behavior of the exact solution. As $U$ is increased, a great deviation shows up between them. At half filling behavior of single-particle Green function is unchanged versus $U$, while in the slave-spin mean-field theory Green function shows ill-defined behavior. Notably, even if particle-hole symmetry is restored in this case, mean-field approximation is effective merely at a small $U$ regime. With a large $U$, although slave spin on mean-field level can describe static properties of ground state of half-filling two-site model exactly, the deviation caused by unphysical states will be magnified with time evolution. This assumption is confirmed again in the non-half-filled situation, where slave-spin method with gauge constraints dose reproduce the desired behavior as analytical solution with all scope of $U$. Our paper provides some insights into the issue of how to improve the accuracy of slave spin, and the hinge is projecting the enlarged Hilbert space onto the physical one.

\begin{figure}
\centering
\includegraphics[width=3.5in]{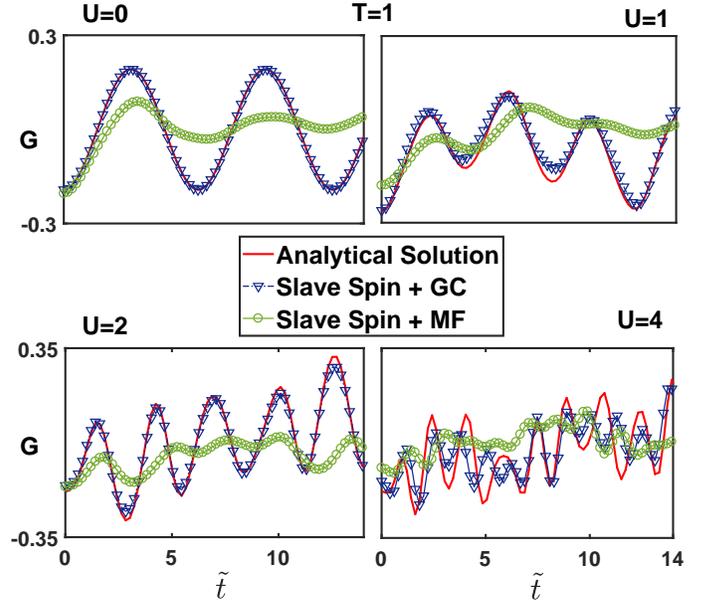}
\caption{The single-particle Green function $G(\tilde{t})$ with evolution of time in finite-temperature ($kT/t=1$) non-half-filled situation. Exact diagonalization (red line), the $\mathbb{Z}_2$ slave-spin MF method (green circle) and the slave-spin method with GC (blue triangle). The results are calculated with different $U$.}
\label{FIG7}
\end{figure}

Here we also provide the results with finite temperature in non-half-filled situation (see in Fig.\ref{FIG6} and Fig.\ref{FIG7}). At finite temperatures, it is necessary to include time evolution of all excited states on average level, weighted by the Boltzmann factor (We take $k=1$ throughout this paper). It demonstrates that the availability of slave spin with mean-field approximation is merely limited to a small parameter regime, where both correlation strength and temperature are small enough compared with hopping integral term. As shown in Fig.\ref{FIG7}, when $kT/t=1$, slave-spin mean-field method accesses to an incorrect behavior of dynamical properties. This is caused by unphysical states which could not be removed completely in mean-field approximation. As we know, $\lambda$ of Lagrange multiplier is dependent on $T$ and $U$. In a high temperature context, thermal fluctuation is quite severe and this two-site system is instable. This instability is also reflected on $\lambda$, where a tiny deviation of $\lambda$ will lead to a great distinction in Hamiltonian and destroy gauge constraints completely. Because of the difficulty in calculating the accurate value of $\lambda$, it is impossible to satisfy constraint conditions and access the exact behavior on mean-field level.

The good news is that only if we could take gauge constraints into account exactly at both two sites, slave-spin method behaves credible at most parameter regime. Only tiny deviation shows up at high temperature together with strong correlation situation (see in Fig.\ref{FIG7}). It turns out that in general situation, slave-spin method could also work well in calculating dynamical quantities as long as we can project the enlarged Hilbert space into the original physical one.

\subsection{\label{sec4:level1D}thermal dynamic properties: heat capacity}

\begin{figure}[h]
\centering
\includegraphics[width=3.5in]{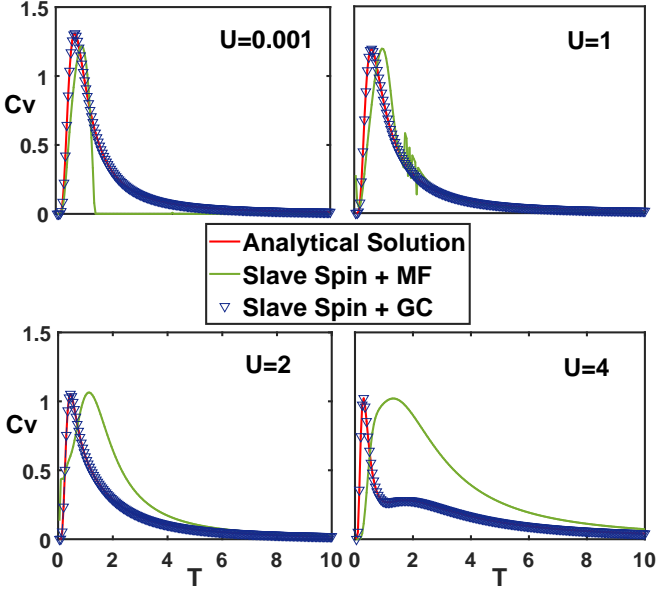}
\caption{The heat capacity $C_v$ in half-filled situation. Exact diagonalization (full red line), the $\mathbb{Z}_2$ slave-spin MF method (full green line) and the slave-spin method with GC (blue triangle). The results are calculated with different $U$.}
\label{FIG8}
\end{figure}

 \begin{figure}[h]
\centering
\includegraphics[width=3.5in]{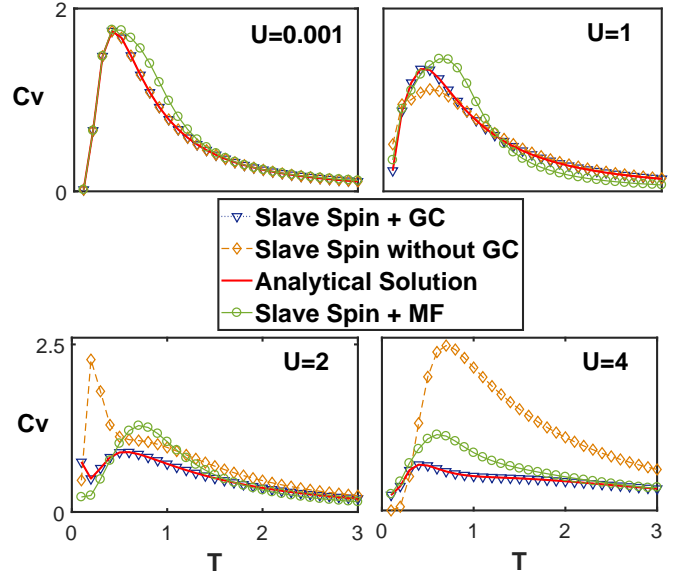}
\caption{The heat capacity $C_v$ in non-half-filled situation. Exact diagonalization (full red line), the $\mathbb{Z}_2$ slave-spin MF method (green circle), the slave-spin method with GC (blue triangle) and also the slave-spin method without GC (yellow diamond).  The results are calculated with different $U$.}
\label{FIG9}
\end{figure}
A stringent test for slave-spin method is to compare its thermal dynamic properties. To benchmark the capacity of slave spin in studying thermal dynamic properties, next we focus on the behavior of heat capacity with all three methods. We derive heat capacity with the aid of partition function, which is obtained by summing up all configurations of this system. As a canonical ensemble, $\mu=0$ so that partition function reads as
\begin{eqnarray}
\begin{split}
\textit{Z}=\sum_ie^{-\frac{E_i}{kT}},
\label{equ51}
\end{split}
\end{eqnarray}
where $E_i$ stands for the eigen-energy on $i-th$ level and $k$ is the Boltzmann constant. Since in Sec.\ref{sec3:level1} we have introduced how do different three methods access to the energy level distribution, with partition function it is easy to access to average value of energy $\langle E\rangle$ and square of energy $\langle E^2\rangle$. Heat capacity is written as
\begin{eqnarray}
\begin{split}
C_v=\frac{\langle E^2\rangle-\langle E\rangle^2}{kT^2}.
\label{equ51}
\end{split}
\end{eqnarray}
As shown in Fig.\ref{FIG8} and Fig.\ref{FIG9}, we demonstrate both half-filled and non-half-filled situation. Notably, in both situation, slave-spin method with exact gauge constraints perfectly reproduce the behavior of analytical solution, while on mean-field level the results could not catch the nature of thermal dynamic properties unless correlation strength is small enough ($U/t<1$). In Fig.\ref{FIG9} we also demonstrate the result without constraint condition at all. Similar with the result of mean-field approximation, in this non-constraint situation the heat capacity calculated with mixed physical and unphysical states is well-behaved merely at small $U$ regime, because the distribution of eigen-energy in slave-spin representation closely resemble the original physical one. These results suggest the great influenced caused by unphysical states, which could be eliminated completely with gauge constraints. The unit of axis is set as $\frac{t}{k}$, which means slave spin with gauge constraints could work well in a large temperature regime.

\section{\label{sec5:level1A}conclusion}
In this paper, we have studied the performance of $\mathbb{Z}_2$ slave-spin method in describing various properties. In order to investigate the accuracy of slave spin, we choose the minimal formulation of previous slave-particle representations and the simplest correlated system, i.e., two-site Hubbard model. Because of the simplicity of formulation and the finite Hilbert space, we directly access to analytical solution and set it as benchmark. By recovering the static, dynamical and thermal dynamic properties with slave-spin method both with exact gauge constraints and mean-field approximation, we can eventually make a fair evaluation about slave spin.

 \par Surprisingly, in half-filled situation slave spin on a mean-field level can perform exactly as well as analytical solution when investigating the static properties of ground state. It turns out that the accuracy of mean-field approximation is ensured by the particle-hole symmetry. However, in more general non-half-filled situation or when we turn to the dynamical properties and thermal dynamic properties, slave-spin method merely work well at the regime where both $U$ and $T$ are small ($U/t<1$ and $kT/t<1$). Although slave-spin mean-field approach is not available in some parameter regime, we could still use it accessing to accuracy behavior of all properties with gauge constraints. Considering constraint condition exactly, only tiny deviation shows up when both $T$ and $U$ is extremely large. All these deviation are caused by unphysical states. At a small $U$ and low temperature situation the influence of unphysical states is tiny, but it would be magnified with strong correlation and strong energy fluctuation.

Our work highlights the importance of identification of physical states in slave-particle method. As long as unphysical states are excluded completely, slave-particle mean-field method could access to exact properties. However, most strongly correlated systems are much more complex and it is difficult to take constraint condition into account exactly, i.e., some unphysical states are unavoidably remaining. For all this, our work finds a way out for slave particle. In some models with specific symmetry, the gauge constraint does not change the formulation of partition functions and thus unphysical states may not impact the expected value of physical observables. As we know, the particle-hole symmetry has successfully simplified slave-particle mean-field theory in many cases. In the future research, we could apply slave particle into this sort of models, such as periodic Anderson impurity model. Besides, some other symmetries may also work in other contexts.
 We hope our work can enlighten slave-particle method to be utilized into suitable area and motivate further investigations. 

\begin{acknowledgments}
We thank Ji-ze Zhao for careful reading and useful suggestions. This research was supported in part by NSFC under Grant No.~11674139,
No.~11704166, No.~11834005, the Fundamental Research Funds for the Central Universities, and PCSIRT (Grant No.~IRT-16R35).
\end{acknowledgments}


\author{Ann Author}
\author{Second Author}%
\affiliation{%
 Authors' institution and/or address\\
 This line break forced with \textbackslash\textbackslash
}%

\collaboration{MUSO Collaboration}

\author{Charlie Author}
\affiliation{
 Second institution and/or address\\
 This line break forced
}%
\affiliation{
 Third institution, the second for Charlie Author
}%
\author{Delta Author}
\affiliation{%
 Authors' institution and/or address\\
 This line break forced with \textbackslash\textbackslash
}%

\collaboration{CLEO Collaboration}

\date{\today}

\nocite{*}

\bibliography{apssamp}

\end{document}